\documentclass[runningheads,a4paper]{llncs}

\usepackage[T1]{fontenc}
\usepackage[ansinew]{inputenc}
\usepackage[english]{babel}
\usepackage{graphicx}
\usepackage{stmaryrd}
\usepackage{amssymb}
\usepackage{color}
\usepackage{listings}
\usepackage{url}
\usepackage{booktabs}
\usepackage{amsfonts}
\usepackage{graphicx}
\usepackage{stmaryrd}
\usepackage{listings}
\usepackage[rgb,dvipsnames]{xcolor}
\usepackage{setspace}
\usepackage{fancyhdr}
\usepackage{subfigure}
\usepackage{float}
\usepackage{epstopdf}
\begin{document}

%\graphicspath{{Figures/}}

\mainmatter  % start of an individual contribution

 %first the title is needed
\title{Anonymously Analyzing Clinical Datasets}
%%
%%% a short form should be given in case it is too long for the running head
\titlerunning{Anonymously Analyzing Clinical Datasets}
%%
%%% the name(s) of the author(s) follow(s) next
%%%
%%% NB: Chinese authors should write their first names(s) in front of
%%% their surnames. This ensures that the names appear correctly in
%%% the running heads and the author index.
%%%
%\author{Nafees Qamar\inst{1}, Yilong Yang\inst{2}, Andras Nadas\inst{1},\\ Zhiming Liu\inst{3}, and Janos Sztipanovits\inst{1}
%}
\authorrunning{N. Qamar, Y. Yang, A. Nadas, Z. Liu, and J. Sztipanovits}
%\institute{Institute for Software Integrated Systems\\
%          Vanderbilt University\\
%          \and
%          University of Macao
%          \and
%          Centre for Software Engineering\\
%          Birmingham City University\\
%           \email{\{nqamar, nadand, sztipaj@isis.vanderbilt.edu\inst{1}, \\yylonly@acm.org\inst{2}, zhiming.liu@bcu.ac.uk\inst{3}\}}
%             }

\author{
Nafees Qamar\inst{1} \and
Yilong Yang\inst{2} \and
Andras Nadas\inst{1}\and\\
Zhiming Liu\inst{3} \and
Janos Sztipanovits\inst{1}}
\institute{Institute for Software Integrated Systems\\
          Vanderbilt University\\
           \email{\{nqamar, nadand, sztipaj\}@isis.vanderbilt.edu}
           \and
           University of Macao \\
           \email{\{yylonly\}@acm.org}
           \and
           Birmingham City University\\
           \email{\{zhiming.liu\}@bcu.ac.uk}
           }
         \maketitle

\begin{abstract}

%Managing clinical data not only requires safeguarding patient's information
%but disseminating and sharing it for clinical research and other secondary
% purposes such as health system planning.
% , which employ statistical methods
%supported by suppression and/or generalization of information.
%Despite the fact that de-identified datasets have is well-documented usage,
%Thus,
 %de-identified datasets end up providing no guarantee for anonymity.

This paper takes on the problem of automatically identifying clinically-relevant patterns in medical datasets without compromising patient privacy. To achieve this goal, we treat datasets as a \emph{black box} for both internal and external users of data that lets us handle
clinical data queries directly and far more efficiently. The novelty of the approach lies in avoiding the data \emph{de-identification} process often used as a means of preserving patient privacy. The implemented toolkit combines
 software engineering technologies such as Java EE and RESTful web services,
to allow exchanging medical data in an unidentifiable XML format as well as
restricting users to the \emph{need-to-know} principle. Our technique also inhibits retrospective processing of data, such as attacks by an adversary on a medical dataset using advanced computational methods to reveal Protected Health Information (PHI). 
The approach is validated on an endoscopic reporting application based on openEHR and MST standards.
From the usability perspective, the approach can be used to query datasets by clinical researchers, governmental or non-governmental  organizations in monitoring health care services to improve quality of care.

%\keywords{Clinical datasets, de-identification, software engineering tools, Java EE, service-oriented architecture, role-base access control}

\end{abstract}

\section{Introduction}

Patients' Electronic Health Records (EHRs) are stored, processed,
and transmitted across several healthcare platforms and among clinical researchers for on-line diagnostic
services and other clinical research.
This data dissemination serves as a basis for prevention and diagnosis of a disease and other secondary purposes such as health system planning, public
health surveillance, and generation of anonymized data for testing. However,
exchanging data across organizations is
 a non-trivial task because of the embodied potential for privacy intrusion.
Medical organizations tend to
%, regardless of whether or not acting on behalf of a patient,
have confidential agreements with patients,
which strictly forbid them to disclose any identifiable information of the patients. Health Insurance Portability and Accountability Act (HIPAA) explicitly
states the confidentiality protection on health information
that any sharable EHRs system must legally comply with.
To abide by these strict regulations, data custodians generally use de-identification\footnote{De-identification process is defined as a technology to delete
 or remove the identifiable information such as name, and SSN from the released information, and suppress or
 generalize quasi-identifiers, such as zip code date of birth, to ensure that medical data is not re-identifiable (the
 reverse process of de-identification.)} techniques \cite{cat}\cite{arg}\cite{sdc} so that any identifiable information on
patient's EHR can be suppressed or generalized.

However, in reality, research \cite{simp}
 indicates that 87\% of the population of U.S. can be distinguished by sex, date of birth and zip code. We can define
 quasi-identifiers as the background information about one or more people in the dataset. If an adversary has knowledge
 of these quasi-identifiers, it can possibly recognize an individual and take advantage of his clinical data.
 On the other hand, we can find out most of these quasi-identifiers have statistical meanings in clinical research.
There exists a paradox between reducing the likelihood of disclosure risk and retaining the data quality.
 For instance, if information related to patients' residence was excluded from the EHR, it would disable related clinical partners to catch
the spread of a disease. Thus, strictly filtered data may lead to failure in operations. Conversely, releasing data including patients' entire information including residence,
sex and date of birth would bring a higher disclosure risk.

\begin{figure}[t]%[!htb]
\centering
\includegraphics[width=0.8\columnwidth]{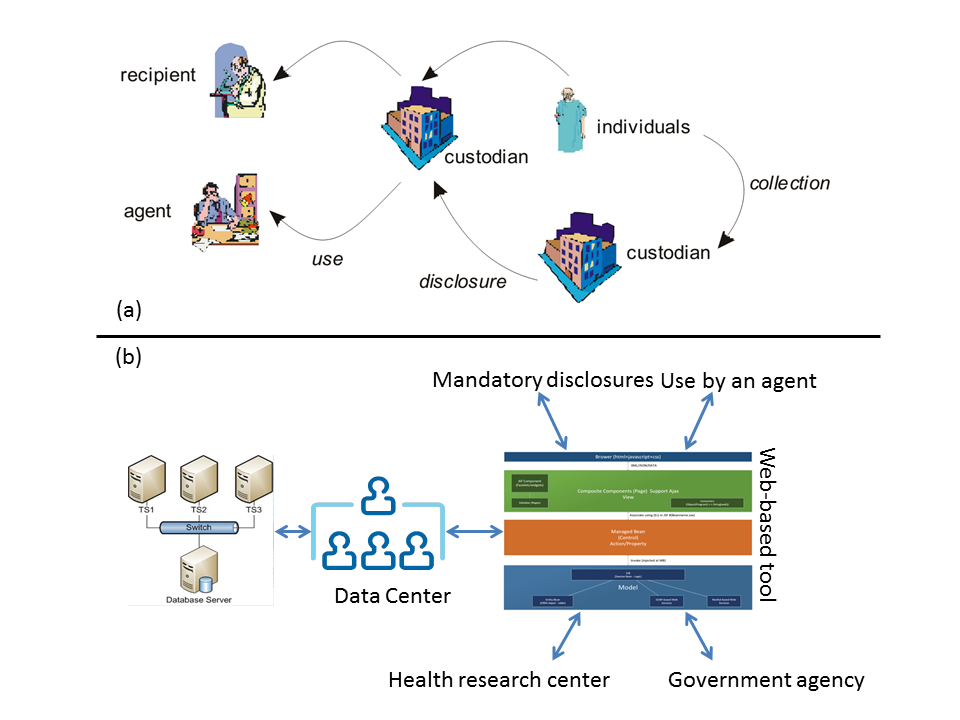}
\caption{(a): shows a traditional lifecycle of medical datasets.
Custodians can be hospitals, agents may be entities working on their behalf, and
recipients are individuals, or organizations such as a pharmaceutical company \cite{el2009overview};
(b) depicts the proposed approach that links external entities to data centers using a web interface. The
approach excludes all direct data accesses on a dataset.}\label{traditionalnewapproach}
%\noindent \dotfill
\end{figure}

In this paper we address the emerging problem of de-identification techniques, namely, the problem of offering de-identified
dataset for a secondary purpose that makes it possible for a prospective user to perform retrospective processing
of medical data endangering patient privacy. Figure ~\ref{traditionalnewapproach} overviews
the proposed technique, and the standard data request process.
 Our approach differs from the traditional techniques in the sense that it employs software engineering
principles
 to isolate and develop key requirements of data custodians and requesters.
We apply Service-Oriented Architecture (SOA) that provides an effective solution for
 connecting business functions across the web---both
between and within enterprises \cite{kreger2001services}. %In this paper we take on a contrary and novel approach to avoid such re-identification risks associated with

We also present a prototype of our evolving toolset, implemented using web services to handle data queries. The results are retrieved in an XML
 data format that excludes all personal information of patients. The basic model used here follows the principles of RESTful
 web services by combining three elements: a \emph{URLs repository} for
identifying resources uniquely corresponding to clinical data queries, \emph{service consumers} requesting data,
 and \emph{service producers} as custodians of clinical data. The idea of combining web services with SQL queries is although not new, but it tends to provide a technological approach to avoid medical data re-identification risks. The implemented toolkit uses Java EE that offers an easy way to develop applications using EJBs. Needless to mention that Java EE is widespread and is largely used by community.

 Our proof-of-concept implementation uses GastrOS, an openEHR \cite{openEHR} database\footnote{\url{http://gastros.codeplex.com}} describing an endoscopic application. The underlying technique
 provides the ability to construct or use stored queries on a clinical dataset.
 Employing this clinical toy data warehouse of the GastrOS prototype is a useful way to demonstrate queries on
 medical data for secondary use. The proposed technique avoids compromising patients' personal information without utilizing de-identification framework tools.
 For instance, the following query can be posed to GastrOS database using our toolkit:

% Number of patients who are still susceptible to developing a Hepatitis B infection even after
%full compliance to the Hepatitis B vaccination schedule ?i.e. The baseline and second detection dates for the HBsAg and Anti-HBs tests turn out to be both negative results.

  \begin{itemize}
 \item [--] \textsf{Find the number of patients who are still susceptible to
 developing a Hepatitis B infection even after full compliance
to the Hepatitis B vaccination schedule--i.e. the baseline and second detection dates for the HBsAg and Anti-HBs
tests both show negative results.}
 \end{itemize}

The set of clinical data queries described in the paper have been crafted with the help of clinical researchers at Vanderbilt University.
Supporting such complex queries
required developing a set of tools, to which this paper provides the first attempt. In contrast to recent developments on big data, this paper does not focus on the management challenges of medical dataset repositories, but rather focuses on software engineering solutions to deal with the challenges of querying medical data endangering patient privacy.
Our approach mainly contributes to the development of privacy preserving techniques on patient data
by treating datasets as \emph{blackbox}. In this way,
disclosure risks associated with patient data are minimized.
One of the key constraints before accomplishing this goal requires keeping the \emph{computability} with data custodians.
Relocating datasets is not only unsafe but leads to data re-identification attempts. To ensure that legitimate users access
and execute clinical data queries, we implement an authentication and authorization mechanism using
role-based access control (RBAC). RBAC offers a flexible architecture that manages
users from different organizations by assigning roles and their corresponding permissions.

The paper proceeds as follows: Section~\ref{related work}
describes the related work; Section ~\ref{gastroscasestudy} states an application
example; Section~\ref{proposedappraoch} presents the technical details of our
approach;  Section~\ref{clinicalqueriessoa} overviews the clinical data queries corresponding to the GastrOS dataset;
Section~\ref{accesscontrol} discusses the authentication and authorization mechanism
connecting users to clinical datasets;  Section~\ref{concandfuturework} summarizes
the work and details some future research directions.

\section{Related Work}\label{related work}

%Demand for de-identified medical datasets is generally increasing, primarily because of the need for
%data collection for identifying clinically-relevant patterns and other estimations by private and governmental
% sectors that account for healthcare quality.

In contrast to some of the existing techniques \cite{DBLP:journals/jamia/OsterLHEMPKSCSFS08} \cite{DBLP:journals/jib/ChoiMBBESSJ07}
\cite{DBLP:conf/amia/McDonaldBDSHADMPD06} \cite{info:doi/10.2196/medinform.2519} \cite{jensweber}, our approach relies on advanced software engineering principles and technologies for analyzing clinical
datasets. For example, caGrid 1.0 \cite{DBLP:journals/jamia/OsterLHEMPKSCSFS08} (now caGrid 2.0), released in 2006, is an
 approach that discusses a complex technical infrastructure for biomedical
research through an interconnected network.
It aims provide support for discovery, characterization, integrated access, and management
of diverse and disparate collections of information sources, analysis methods, and applications in
biomedical research. caGrid 1.0 has been initially designed only for cancer research. caGrid combines
Grid computing technologies and the Web Services Resource Framework (WSRF) standards to provide
a set of core services, toolkits for the development and deployment of new community provided services,
and APIs for building client applications. However, caGrid does not focus on an explicit query mechanism to infer details from medical datasets, as the one proposed here.
Similar work in \cite{DBLP:journals/jib/ChoiMBBESSJ07} discusses a combined interpretation of
biological data from various sources. This work, however, considers the problem of continuous
updates of both the structure and content of a database and proposes the novel database SYSTOMONAS
for SYSTems biology of pseudOMONAS. Interestingly, this technique combines a data warehouse concept
with web services. The data warehouse is supported by traditional ETL  (extract, transform, and load) processes
and is available at \url{http://www.systomonas.de}.

De-identification techniques for medical data have been studied and developed
 by statisticians dealing with integrity and confidentiality issues of statistical data.
The major techniques used for data de-identification are (i) CAT (Cornell Anonymization Kit) \cite{cat}, (ii) $\mu$-Argus \cite{arg}, and (iii) sdcMicro \cite{sdc}. CAT anonymizes
 data using generalization, which is proposed \cite{gen} as a method that specifically replaces values
 of quasi-identifiers into value ranges. $\mu$-Argus is an acronym for Anti-Re-identification General Utility
 System and is based on a view of safe and unsafe microdata that is used at Statistics Netherlands,
 which means the rules it applies to protect data comes from practice rather than the precise form of rules.
 Developed by Statistics Austria, sdcMicro is an extensive system for statistical computing.
% sdcMicro is developed by Statistics Austria based as a highly extensive system for statistical computing.
 Like $\mu$-Argus, this tool implements several anonymization methods considering different types of variables.
 We have reported \cite{DBLP:conf/fhies/LiuQQ13} a comparison on the efficacy of these numerical methods that
 are used to anonymize quasi-identifiers in order to avoid disclosing individual's sensitive information. The
 Privacy Analytics Risk Assessment Tool (PARAT) \footnote{\url{http://www.privacyanalytics.ca/software/}} is the
 only commercial product available so far for de-identifying medical data.
 Our quantitative analysis  \cite{DBLP:conf/fhies/LiuQQ13} of de-identification tools shows that
de-identifying data provides
 no guarantee of anonymity \cite{DBLP:journals/tkde/Samarati01}.  A study \cite{DBLP:journals/jamia/BenitezM10} also shows that organizations
 using data de-identification are vulnerable to re-identification at different rates.

Another approach \cite{DBLP:conf/amia/McDonaldBDSHADMPD06} describes a special query tool developed for the Indianapolis/Regenstrief Shared Pathology
Informatics Network (SPIN) and integrated into the Indiana Network for Patient care (INPC). This tool allows
retrieving de-identified data sets using complex logic and auto-coded final diagnoses, and it
supports multiple types of statistical analyses. However, much of the technical details have not been published; for example,
the use of complex logic. This and  other similar efforts \cite{DBLP:conf/amia/PratherLGHHH97}
 are mostly database-centric. A slightly similar work to this paper has been developed
 at Massachusetts General Hospital (QPID Inc., \footnote{\url{http://www.qpidhealth.com}}),
offering solutions at a commercial level, but no prototype is available to experiment with. A Web-based approach for enriching the capabilities of the data-querying system is also
developed \cite{info:doi/10.2196/medinform.2519} that considers three important aspects including the interface
design used for query formulation, the representation of
query results, and the models employed for formulating query criteria. The notion of differential privacy \cite{DBLP:conf/icalp/Dwork06}
 aims to provide means to maximize the accuracy of queries from statistical databases while minimizing the chances of identifying its records.

Our analysis shows that the effort to secure medical datasets is mainly two-faceted: 1) most research endeavors have explored the design and
 development of de-identifi\-cation tools, and, 2) some work, mostly led by medical doctors, has
  tried to address the construction of clinical queries, but
 they do not provide technical details on the construction of their toolsets. Our approach that treats medical datasets as blackbox
mainly considers the automation of services expected from a data custodian in order to minimize data disclosure risks and
 making clinical datasets easily accessible for internal and external users.

\section{GastrOS: An Example Application}\label{gastroscasestudy}

GastrOS\footnote{\url{http://gastros.codeplex.com}}, an openEHR database describing an
 endoscopic application, is used as a case-study of electronic medical data.
 This application formed part of the research
done at University of Auckland by Koray Atlag in 2010 that investigated software maintainability and
interoperability. For this, the domain knowledge model of Archetypes and
 Templates of openEHR has driven the generation of its graphical user interface.  Moreover, the data
 content depicting the employed terminology, record structure and semantics were based on the
 Minimal Standard Terminology for Digestive Endoscopy (MST) specified by the World
 Organization of Digestive Endoscopy (OMED) as its official standard.

Employing the clinical toy data warehouse of the GastrOS prototype is a useful way to demonstrate
clinical research based queries on medical data for secondary use without compromising patients' personal information
by using the approach proposed here.
The queries shown here focus on endoscopic findings that provide valuable anonymized information to clinicians.
The implemented queries are to be mainly used by medical practitioners and health decision-makers alike to help them in their clinical
  management of patients at the point-of-care and in formulating appropriate health policies, respectively. For example, the following queries are obtained through brainstorming
with medical doctors to illustrate our approach.

\begin{itemize}
\item[--] \textsf{Total number of dialysis endoscopic examination from January 1, 2010 to December 31, 2010.}
\item[--] \textsf{Top 5 diagnoses for those patients who received endoscopic examination and the
number of cases for each diagnosis from January 1, 2010 to December 31, 2010.}
\item[--] \textsf{Age profile of endoscopic patients from January 1, 2010 to December 31, 2010 ? i.e. number of dialysis patients
belonging to each of the age bracket [below 18; 18 to below 40; 40 to below 60; 60 and above.}
\item[--] \textsf{Number of patients who are still susceptible to developing a Hepatitis B infection even after
full compliance to the Hepatitis B vaccination schedule?--i.e., the baseline and second detection dates for the HBsAg and Anti-HBs tests both show negative results.}
\end{itemize}

%\begin{itemize}
%\item[--] \textsf{Total number of dialysis treatment done from January 1, 2010 to December 31, 2010.}
%\item[--] \textsf{Top 5 diagnoses for those patients who received dialysis treatment and the number of cases
%for each diagnosis from January 1, 2010 to December 31, 2010.}
%\item[--] \textsf{Age profile of dialysis patients from January 1, 2010 to December 31, 2010--i.e., Number of
%dialysis patients belonging to each of the age bracket [below 18; 18 to below 40; 40 to below 60; 60 and above].}
%\item[--] \textsf{Number of patients who are still susceptible to developing a Hepatitis B infection even after full compliance
%to the Hepatitis B vaccination schedule--i.e., The baseline and second detection dates for the HBsAg and Anti-HBs
%tests turn out to be both negative results.}
%\end{itemize}

\noindent
The queries given above are only a subset of original queries. The
database structure of GastrOS application is described below.

\subsection{GastrOS data structure}

\begin{figure}[!htb]
  \centering
  \includegraphics[width=1\textwidth]{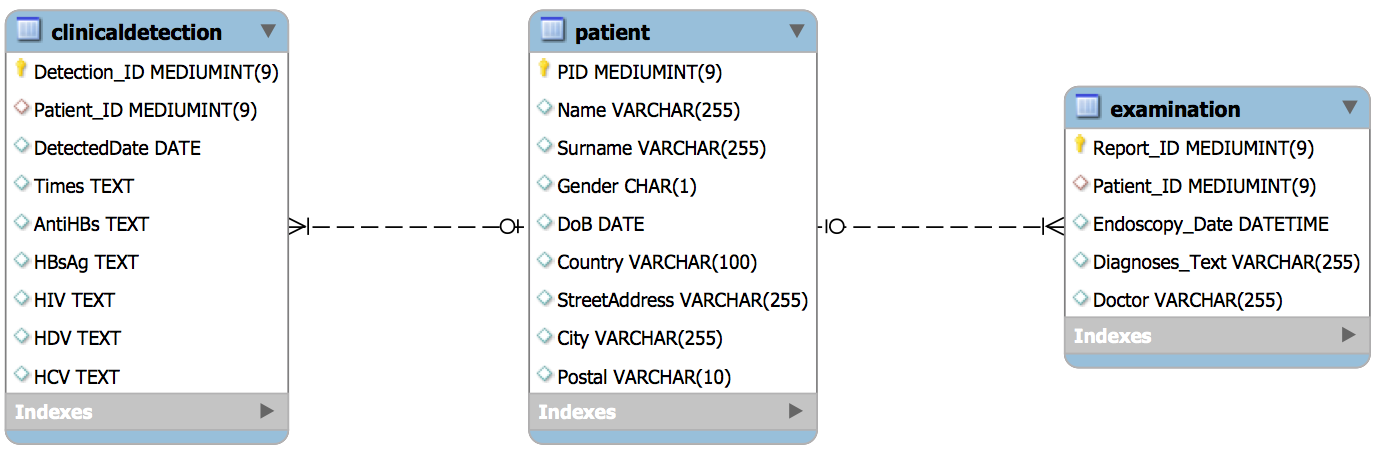}
  \caption{E-R diagram}\label{ERaa}
\end{figure}

Figure \ref{ERaa} describes the data structure of the GastrOS database.
GastrOS database contains the following tables: the \textsf{clinicaldetection} (doctor detection records),
\textsf{patient} (patient information), and \textsf{examination} (examination records)
 tables are stored in the database.

\begin{figure}[!htb]
  \centering
  \includegraphics[width=\columnwidth]{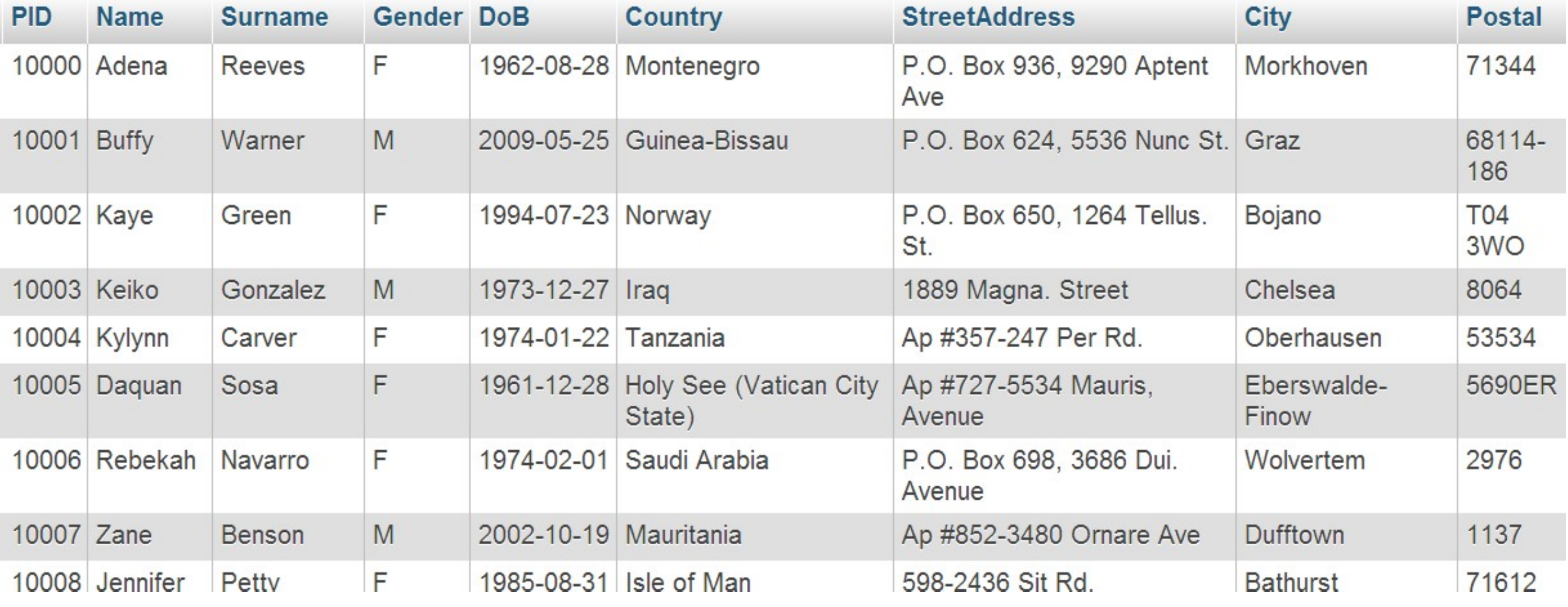}\\
  \caption{Data generated of patient table}\label{Patient}
\end{figure}

%We extend  All the data we used are stored in the database, we extend GastrOS`s database with
% access control. The E-R diagram show in Figure \ref{ER}. There are two parts in the database,
%
The \textsf{patient} table has two relations: one patient may have more than one clinical detection record
 or examination record by doctor(s), so the \textsf{patient id} is added as a foreign key in tables \textsf{ClinicalDection} and \textsf{Examination}.
GastrOS is a toy database example with insufficient amount of data available. The original database contains less than 20 rows in each table
that makes is not useful for our SQL queries. Therefore, we automatically generated virtual data of 10,000 entries
 (note that any real data on patients also cannot be published.) An example of the generated data is given in Figure ~\ref{Patient}.
Table ~\ref{DataGenerationa} provides the up-to-date information on the number of entries in each column of the GastrOS
database.

\begin{table}
  \centering
  \footnotesize
   \caption{Generated data in tables}\label{DataGenerationa}
      \begin{tabular}{lll}
      \toprule
      Table  & Row  & Size \\
      \midrule
      ClinicalDetection & 6,393 & 432 KB \\
      Examination & 2,020 & 272 KB \\
      Patient & 1,881 & 224 KB \\
      Sum & 10,294 & 928 KB \\
      \bottomrule
    \end{tabular}
\end{table}
%\begin{figure}[!htb]
%  \centering
%  % Requires \usepackage{graphicx}
%  \includegraphics[width=0.8\textwidth]{clinicaldetection.jpg}\\
%  \caption{Clinical Detection}\label{clinicaldetection}
%\end{figure}
%\begin{figure}[!htb]
%  \centering
%  % Requires \usepackage{graphicx}
%  \includegraphics[width=0.8\textwidth]{examination.jpg}\\
%  \caption{Examination}\label{examination}
%\end{figure}

\section{The Proposed Technique}\label{proposedappraoch}

The proposed approach implements a three-tier application and is devoid of releasing medical datasets, as opposed
to traditional techniques. The major purpose and characteristic of the technique extends relatively new software technologies
for supporting clinical data queries.
%To elaboa the toolkit we mainly choose to describe here the technical details. Figure ~\ref{Architecture} describes the architecture of the tool used to produce XML-based clinical data corresponding to the queries.
In order to support clinical queries under consideration, we develop an integrated application using SOA and Java EE (Enterprise Edition), to extract data from GastrOS database.
There are a plenty of other commercial containers such as JBOSS (Redhat), Websphere (IBM), Weblogic and Glassfish (Oracle), which could be used for our purpose. However, our prototype tool combines Java EE based on JSF Primeface, EJB,
and Java Persistence Architecture API (JPA). JPA is a Java specification for accessing,
persisting, and managing data between Java objects / classes and a relational database. REST architecture, underlying
RESTful web services, treats everything as a resource and is identified by an URI. Resources are handled using
POST, GET, PUT, DELETE operations that are identical to Create, Read, Update and Delete (CRUD) operations. Note that in our toolkit it is suffice to implement \textsf{Read} operations for handling the described queries.
Every request from a client is handled independently, and it must contain all the required information to interpret the request.

\begin{figure}[!htb]
  \centering
  \includegraphics[width=\columnwidth]{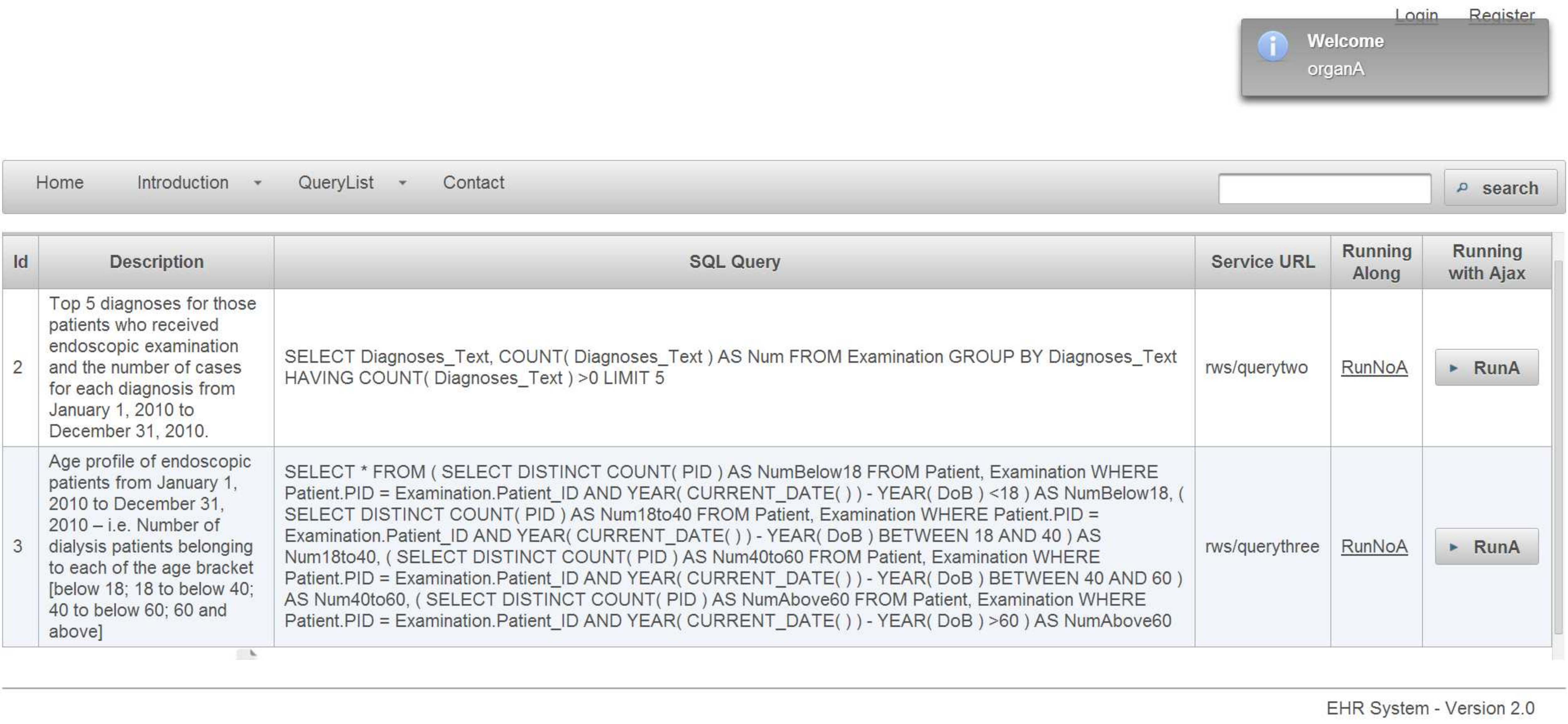}\\
  \caption{The list of authorized roles for the Organization A}\label{OrganA}
\end{figure}

%[language=JAVA,caption={Java code of web service},label={javacode}]
\begin{verbatim}
# Code for the Restful-based web service.
@Path("queryone")
public class QueryOne {
    @Context
    private UriInfo context;
    @EJB
    QueryBean bean;
    @GET
    @Produces("application/xml")
    public String getHtml() {
        // TODO return proper representation object
    	
      String sql = "select Country, COUNT(Report_ID ) AS" + 
            "TotalNum " +
			"FROM examination, patient " +
			"WHERE examination.Patient_ID = " + 
            "patient.PID " +
			"AND Endoscopy_Date " +
			"BETWEEN \'2010-1-1\' " +
			"AND \'2010-12-30\' " +
			"GROUP BY Country   " +
			"Order By TotalNum desc ";
      String f =  bean.query(sql);   	
      return f;
    }
}
\\For the method query:
public String query(String sql)
{
    String result = "";
	Query query =  emf.createEntityManager().
         createNativeQuery(sql);	
	@SuppressWarnings("unchecked")
	List<Object[]> list = query.getResultList();
           ......
}
\end{verbatim}

\section{Implementing Clinical Queries using SOA}\label{clinicalqueriessoa}

Web-based authorization and authentication is enforced using role-based access control, before allowing any queries to be accessible
by external entities. For instance, the first two queries are shown in Figure ~\ref{OrganA}. They are linked
to \textsf{Organization A}, that shows a limited access varying according to the enabled permissions by a security administrator. Thereby, execution of the queries is managed by access control features of the tool. Some of the queries and their corresponding data are given below. 
SQL queries, exception results, and running time are presented in columns 1, 2, and 3, respectively of the Figure ~\ref{OrganA}.

Note that
 XML-based format is devoid of platform and programming language dependencies. Using this Web-based approach a diverse set of queries
can be supported to query clinical data repositories. For the RESTful-based web services before executing a query,
it should have a URL stored in database, that is the table \textsf{urlforwebservice.}

%A code snippet is given in Listing ~\ref{javacode} that reveals how the SQL queries are constructed.
Note that all the data saved in a program are objects; nonetheless, our database has actually been represented in the form of relational tables.
For this, it needs to implement some ORM (Object-Relational Mapping) techniques.
In our prototype implementation we have used JPA (Java Persistence API),
because it comes with Java EE technique framework and can be run in either
native SQL, or in an object form to allow data manipulation. For instance, we show a \emph{service} code snippet above.
\textsf{@Path} show the URL address for this web service, \textsf{@GET} is the method of Restful-based web service,
that can be used for other reasons such as \textsf{@UPDATE @DELETE @POST}.
Upon invoking a web service using URL in browser or a session bean, the SQL can be executed
and return result by query method which invokes the entity manager of JPA. Below, we list some sample clinical queries as well as their output in an XML format.

%[language=XML,caption={Generated XML data},label={result2}]
\begin{verbatim}
#Number of patients for each gender who are still susceptible
to developing a Hepatitis B infection even after full compliance
to the Hepatitis B vaccination schedule --i.e. the baseline and 
second detection dates for the HBsAg and Anti-HBs tests both
show negative results.

<dataset>
<item>
<element>F</element>
<element>184</element>
</item>
<item>
<element>M</element>
<element>192</element>
</item>
</dataset>
\end{verbatim}

\begin{verbatim}
# Top 5 diagnoses for those patients who received dialysis 
treatment and the number of cases for each diagnosis from
January 1, 2010 to December 31, 2010.

<?xml version="1.0" encoding="utf-8"?>
<dataset>
 <item>
  <element>
    Diagnoses_Text Colon: Primary malignant tumor,
    Quiescent Crohn's disease
  </element>
  <element>421</element>
 </item>
 <item>
  <element>
    Diagnoses_Text Esophagus: Normal, Ectopic gastric mucosa
  </element>
  <element>394</element>
 </item>
 <item>
  <element>Esophagus: Reflux esophagitis</element>
  <element>414</element>
 </item>
 <item>
  <element>Esophagus: Varices certain</element>
  <element>406</element>
 </item>
 <item>
  <element>Esophagus:Barrett's esophagus</element>
  <element>365</element>
 </item>
</dataset>
\end{verbatim}

\subsection{Enabling dynamic clinical queries}

\begin{figure}[!htb]
  \centering
  \includegraphics[width=\columnwidth]{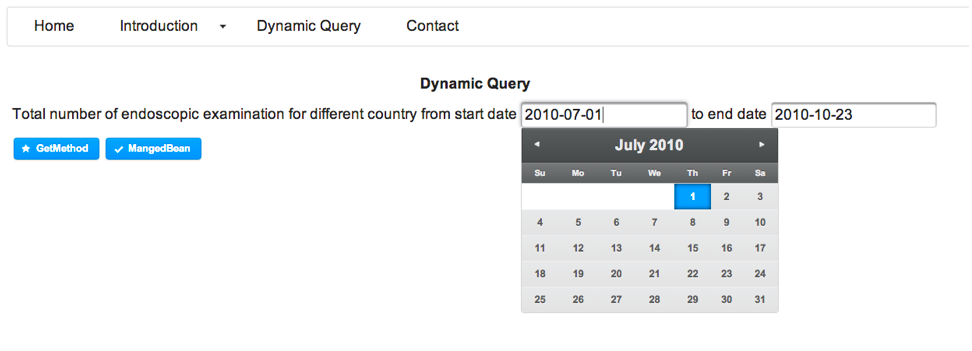}\\
  \caption{Interface for executing runtime clinical queries}\label{dynamicQ}
\end{figure}

\begin{figure}[!htb]
  \centering
  \includegraphics[width=\columnwidth]{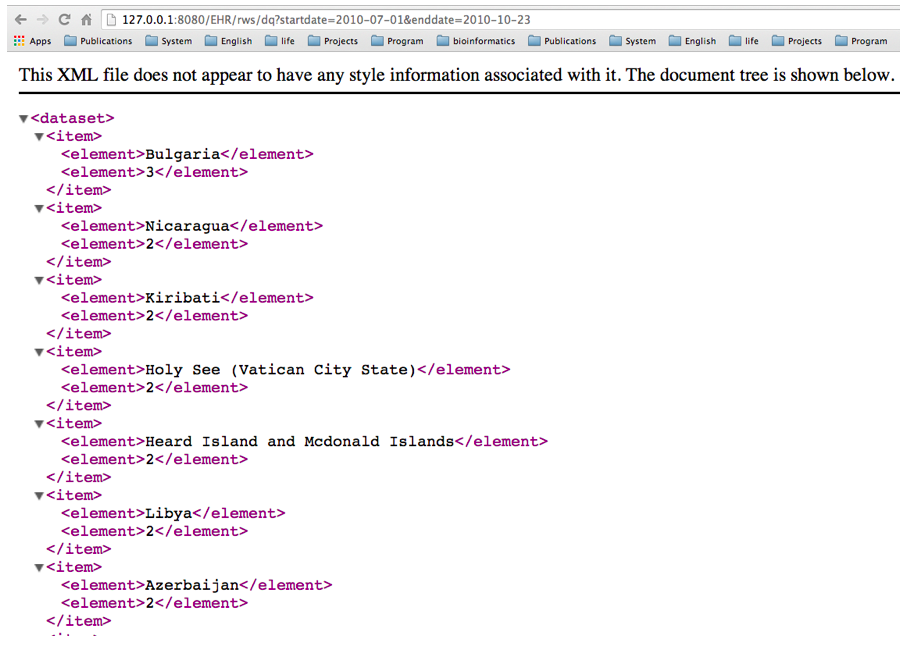}\\
  \caption{The retrieved data in XML format corresponding to the query in ~\ref{dynamicQ}}\label{dynamicQdata}
\end{figure}

The construction and execution of clinical queries on a given dataset are implemented through a web-interface of the tool. The interface allows a user to dynamically construct a clinical query on a dataset. Thus, it adds a greater flexibility to the query mechanism in developing user-oriented analysis of a dataset. For instance, Fig. ~\ref{dynamicQ} demonstrates how to execute a query such as \textsf{"Total number of dialysis endoscopic examination of a country starting and ending on a particular date, respectively."}, followed by the output in Fig ~\ref{dynamicQdata}.

These queries show that all specific details on patients are avoided when executing a query, which also means that it disables all direct accesses to patient records. It is actually realized by providing a more aggregated form of data on patients instead of conventional techniques that provide medial datasets to infer such details. Note that the toolkit does not allow any query that provides specific information on patients, such as \textsf{"Provide details of all patients with a certain age}". These queries are directly irrelevant to researchers since they are mainly interested in collective analysis on a dataset. The idea of combining web services with SQL queries is although not new, but it tends to provide a technological solution to a technological problem avoiding medical data re-identification risks. The rationale Using Java EE stems from the fact that it provides an easy way to develop applications, for example, EJB are convenient to use by adding only one annotation. Java EE is also widespread being largely used both in academia and industry.

%Then, in 2009, JEE 6 was released. Development is so easy. Finally! For example, you have to add only one annotation and your EJB is ready! Of course, the developers of the Spring framework did not sleep. Much new stuff was added. Today, you can create a Spring application without any one XML file as I have read in a ?No Fluff Just Stuff? article some weeks ago. Besides, several really cool frameworks were added to the Spring stack, e.g. Spring Integration, Spring Batch or Spring Roo.
%
%Today (November, 2011), both JEE and Spring are very widespread and have a large community. Much information is available for both, e.g. books, blogs, tutorials, etc.

\section{Authentication and Authorization Process}\label{accesscontrol}

Our toolset implements the Role-based Access Control (RBAC) \cite{DBLP:journals/tissec/FerraioloSGKC01} \cite{DBLP:conf/fhies/QamarFLL12} \cite{DBLP:conf/icfem/QamarLI11}. RBAC provides a suitable mechanism to restrict user's access on resources, such as to perform 
operations including insert, delete, append, and update on a medical dataset. The data model of RBAC is
based on five data types: users, roles, objects, permissions
and executable operations by users on objects.

\begin{figure}[!htb]
  \centering
  \includegraphics[width=\columnwidth]{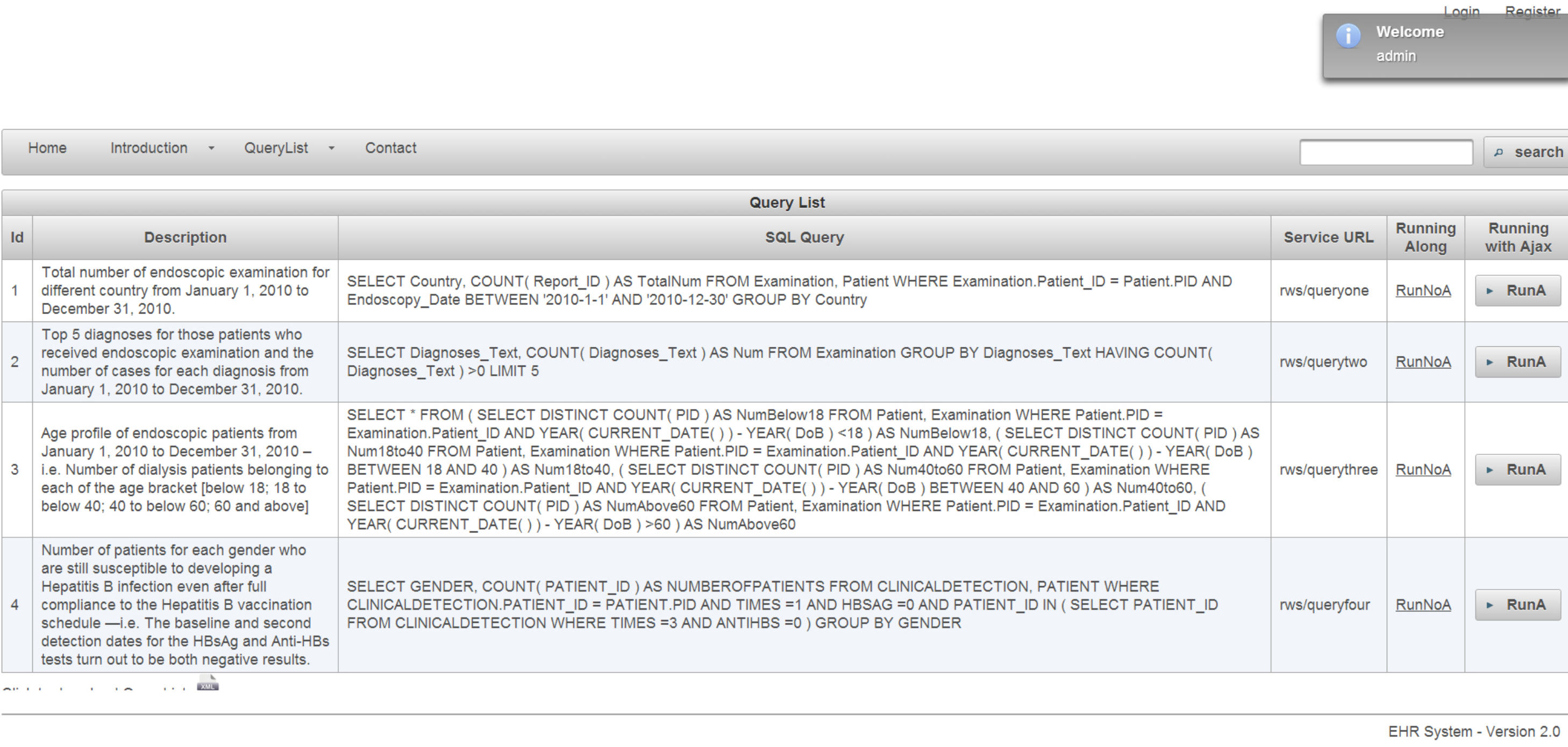}\\
  \caption{Query list for role of administrator}\label{admin}
\end{figure}

\begin{figure}[!htb]
  \centering
  \includegraphics[width=\columnwidth]{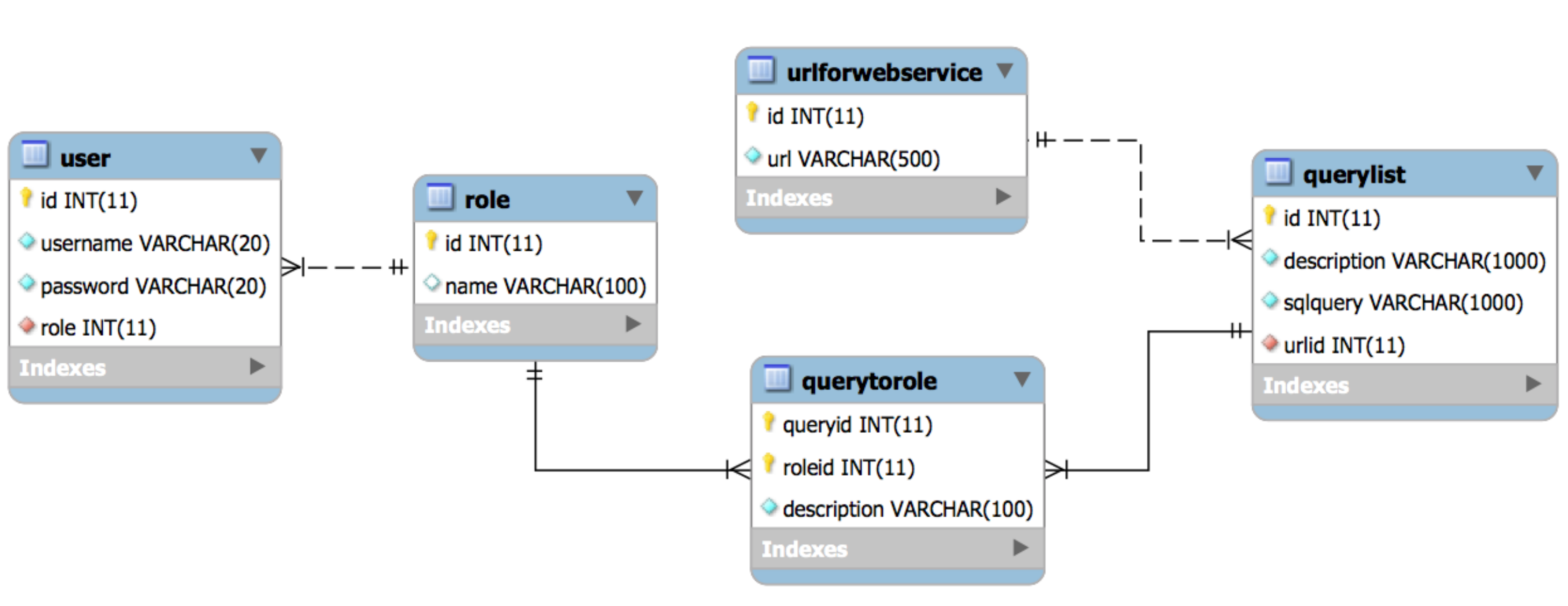}\\
  \caption{E-R diagram}\label{ER}
\end{figure}

\noindent
A sixth data type, session, is used to associate roles temporarily to users.
A role is considered a permanent position in an organization whereas a given user can be switched with another user for that role.
Thus, rights are offered to roles instead of users. Roles are assigned to permissions that can later be exercised by users playing these roles.
Modeled objects in RBAC are potential resources to protect. Operations are viewed as application-specific user functions.
%UA is user assignment and PA is permission assignment.
For example, Fig. ~\ref{admin} shows a list of queries provided to an administrator role.

%\begin{itemize}
%
%	\item [--] $UA \subseteq USERS \times ROLES$, a many-to-many mapping between users and roles; UA specifies which roles can be played by a given user;
%
%	\item [--] $PA \subseteq PRMS \times ROLES$, a many-to-many mapping permission-to-role; PA expresses which roles may be granted a given permission;
%
%	\item [--] $user\_sessions (u:USERS) \rightarrow 2 ^{SESSIONS}$, the mapping of user \textit{u} onto a set of sessions;  it lists the current sessions of a given user;
%
%%	\item [--] $session\_roles (s:SESSIONS) \rightarrow 2^{ROLES}$, the mapping of session \textit{s} onto a set of roles; it lists the current roles of a given user in a given session;
%%
%%	\item [--] $PRMS = 2^{(OPS \times OBS)}$, the set of permissions. Permissions are regarded as an approval to perform operations on RBAC protected objects.
%
%\end{itemize}

To maintain a set of permissions on GastrOS database, we use the constructs from RBAC maintain, and enlist entries in corresponding tables \textsf{users},  \textsf{roles},
 textsf{querytoroles}, \textsf{querylist}, and \textsf{urlforwebservice}.
The database tables include \textsf{user}, \textsf{role}, \textsf{querylist}, \textsf{querytorole}, and \textsf{urlwebservice}.
We create a user account in user table with the assigned role. Here, all the roles are defined in \textsf{role} table.
Users privileges and a list of queries are defined in tables \textsf{querytorole} and \textsf{querylist}, respectively.
URLs are stored in the \textsf{urlwebservice} table. For example, logging in as administrator provides five SQL
queries shown in Figure \ref{admin}, whereas logging in as \textsf{organization A} allows a restricted set of SQL queries as given in Figure \ref{OrganA}.
Security management is supervised by an administrator who can do deletion, addition of roles as required.
Using RBAC allows users to take multiple roles, for example, the user X could
 act as researcher that belongs to organization A, but can be assigned another role from the set of roles. Similarly, a permission can be associated
 to many roles depending on the RBAC policy. The multi-to-multi relation between roles and queries that is given
 in the \textsf{querytorole} table.

% to reach the 3NF (Third Normal Form)

\subsection{Avoiding SQL injections and sensitive information release}

Web application security vulnerabilities occur in cases when an attacker or a authorized user
tries to submit and execute a database SQL command on a web application, and thus, a back-end
database is exposed to an adversary. These SQL injections can be avoided if queries are validated and filtered
before their execution, and are checked against input data or any encoding made by a user. To
 prevent similar security issue in our web application we first authenticate the
 user input against a set of defined rules given below:
\vspace{2mm}

\hspace{10mm} $BlockList = \{name, age, address, zipcode\}$ \\\indent
\hspace{10mm} $Anti\verb|-|injectionList = \{', '', etc.\}$
\vspace{2mm}

%for de-identification \\ for prevent SQL injection.

Note that the special characters given in a block list helps to avoid  SQL injections.
The set \textsf{BlockList} disables all possible access to attributes in a table such as name,
age, address, and zip code to keep the fetched data completely anonymized.
Set members in \textsf{injectionList} filters out three possible vulnerable inputs,
i.e., \textsf{,, etc.} so that any similar attempts could be restricted. Here are
the filters that check inputs against \textsf{BlockList, injectionList}.
Before running a web service, these two atomic services are always invoked
to avoid identifying the actual patients and SQL injections.

\begin{itemize}
  \item Service one: Checks input string for characters in \textsf{BlockList}.
\end{itemize}

\begin{verbatim}
bool CheckDeIdentification(String s)
{
    Check Input string s,
    if it contain character in BlockList,
    return false. Otherwise true.
}
\end{verbatim}
\begin{itemize}
  \item Service two: Checks input string for characters in \textsf{Anti-injectionList}.
\end{itemize}

\begin{verbatim}
bool CheckInjection(String s)
{
    Check Input string s,
    if it contain character in Anti-injectionList,
    return false. Otherwise true.
}
\end{verbatim}

%The prototype system presented here can further be complemented with more security filters
%needed for the construction of dynamic queries.

\section{Conclusions and Future Perspectives}\label{concandfuturework}
%This paper presents a technique for automatic identification of clinically-relevant patterns in medical data.
%Unlike the traditional approach
%that extensively relies on the data \emph{de-identification} process before releasing
%patients' information, our approach excludes this process and handles
%clinical data queries directly. We propose and implement the idea of treating medical
%dataset as a \emph{blackbox} for both internal and external users of data
%for preserving patient privacy. For this, our integrated toolkit combines
%advanced software engineering technologies such as Java EE and RESTful web services,
%which allows to exchange data in an anonymous XML format and
%restricts users to computed information.
%The traditional approach makes it possible for an adversary
%to succeed in data \emph{re-identification} attempts by applying advanced computational techniques.
%The paper addresses this shortcoming by disallowing
%retrospective processing of data.
%We validate our approach on an endoscopic reporting application based on openEHR and MST standards.
%The implemented prototype system
%can be used to query medical data, by clinical researchers, governmental or non-governmental  organizations in monitoring health care services to improve quality of care.

We presented a technique for automatic identification of clinically-relevant patterns in medical data.  The main contribution of
this paper is in defining and presenting an alternative approach to the data de-identification techniques commonly employed for anonymizing clinical datasets.
Our technique treats datasets as \emph{blackbox} and allows data custodians to handle clinical data queries directly.
Relocating a dataset
not only endangers anonymity of patients, it allows adversaries to apply
advanced computational methods for retrospective processing of data. As clinical data is frequently updated,
our approach enables data custodians to provide up-to-date resources to their users. We integrate RESTful web services and Java EE with
a backend clinical database exchanging anonymous XML data, enabling them to be language and technology independent. Java EE, due to equipped with EJBs, is easy to use for developing applications.

In circumstances related to sharing of patients' data,
complex administrative regulations are placed at different levels of management that sometimes unnecessarily
complicate the data acquisition process. Providing a tool support for linking data custodians and data requesters using software engineering techniques could pave
the way to query clinical datasets more transparently and systematically. We explored new ways of anonymously analyzing clinical datasets. Our future work includes expanding the
approach to more complex databases and supporting an enriched interface for analyzing bigger data repositories. We are currently dealing with the challenge of replacing de-identification techniques in use for de-identifying specific attributes in a database table, for example, patient id, and a doctor needing
to find patients who had an increase of systolic blood pressure within a specific period, or patients with steady
states of diastolic blood pressure for more than a week. Our future work considers incorporating such queries into
the toolset, including implementing ETL processes such as in data warehouses
to support clinical data analyses on large-scale integrated databases.

\section{Acknowledgments}

The work presented in this paper was funded through
National Science Foundation (NSF) TRUST (The Team for
Research in Ubiquitous Secure Technology) Science and
Technology Center Grant Number CCF-0424422. Its contents are
solely the responsibility of the authors and do not necessarily
represent the official views of the NSF. 
This work has been partly supported by the project SAFEHR of Macao Science and Technology Development Fund (MSTDF) under grant 018/2011/AI.
%\bibliographystyle{abbrv}
%\bibliography{reference}

\end{document}